\documentclass[journal,twoside,web]{ieeecolor}
\usepackage{generic}
\usepackage{cite}
\usepackage{amsmath,amssymb,amsfonts}
\usepackage{algorithmic}
\usepackage{graphicx}
\usepackage{textcomp}
\usepackage{caption}
\usepackage{subcaption}
\usepackage[hidelinks]{hyperref}
\usepackage{mathtools}
\def\BibTeX{{\rm B\kern-.05em{\sc i\kern-.025em b}\kern-.08em
    T\kern-.1667em\lower.7ex\hbox{E}\kern-.125emX}}
\begin{document}
\title{Analysis of a Deep Learning Model for 12-Lead ECG Classification Reveals Learned Features Similar to Diagnostic Criteria}
\author{Theresa Bender, \IEEEmembership{Student Member, IEEE}, Jacqueline M. Beinecke, Dagmar Krefting, Carolin Müller, Henning Dathe, Tim Seidler, Nicolai Spicher, \IEEEmembership{Member, IEEE}, and Anne-Christin Hauschild
\thanks{Submitted on 14.11.2022.
This research was funded by the German Federal Ministry of Education and Research (grant no. 16TTP073 11, and HiGHmed, grant no. 01ZZ1802B) and the Lower Saxony “Vorab” of the Volkswagen Foundation and the Ministry for Science and Culture of Lower Saxony (grant no. 76211-12-1/21).}
\thanks{T. Bender, J. M. Beinecke, D. Krefting, H. Dathe, N. Spicher and A.-C. Hauschild are with the Department of Medical Informatics, University Medical Center Göttingen, Göttingen, 37075 Germany. (e-mail: theresa.bender@med.uni-goettingen.de).}
\thanks{C. Müller and T. Seidler are with the Department for Cardiology \& Pneumology/Heart Center, University Medical Center Göttingen, Göttingen, 37075 Germany.}
\thanks{\textit{(N. Spicher and A.-C. Hauschild are co-last authors.)}}}

\maketitle

\begin{abstract}
Despite their remarkable performance, deep neural networks remain unadopted in clinical practice, which is considered to be partially due to their lack of explainability. In this work, we apply explainable attribution methods to a pre-trained deep neural network for abnormality classification in 12-lead electrocardiography to open this ”black box” and understand the relationship between model prediction and learned features. We classify data from two public databases (CPSC 2018, PTB-XL) and the attribution methods assign a ”relevance score” to each sample of the classified signals. This allows analyzing what the network learned during training, for which we propose quantitative methods: average relevance scores over a) classes, b) leads, and c) average beats. The analyses of relevance scores for atrial fibrillation and left bundle branch block compared to healthy controls show that their mean values a) increase with higher classification probability and correspond to false classifications when around zero, and b) correspond to clinical recommendations regarding which lead to consider. Furthermore, c) visible P-waves and concordant T-waves result in clearly negative relevance scores in atrial fibrillation and left bundle branch block classification, respectively. Results are similar across both databases despite differences in study population and hardware. In summary, our analysis suggests that the DNN learned features similar to cardiology textbook knowledge.
\end{abstract}

\begin{IEEEkeywords}
atrial fibrillation, electrocardiogram, explainable artificial intelligence, integrated gradients, layer-wise relevance propagation, left bundle branch block
\end{IEEEkeywords}

\section{Introduction}
\label{sec:introduction}
The development and evaluation of algorithms for automatic interpretation of biosignals has attracted great interest in the last decade. Biosignals are time series, i.e. they are ordered sequences of measurements, which are usually acquired in successive and equally-spaced time intervals. Typical examples are the electrocardiogram (ECG) representing the electrical activity of the heart or the electroencephalogram (EEG) representing brain activity. The temporal ordering discriminates biosignals from many other types of biomedical data without any order, such as lab tests or sequencing, and introduces challenges in their interpretation by humans and algorithms alike. Next to measurement artefacts including loss of electrode contact, signals are influenced by other physiological processes, for example ECG by respiration, and (in)voluntary movement of the patient.

Traditionally, the field of ECG signal processing was dominated by methods based on mathematical or physical models recreating human physiology. Human experts defined semantic models or features which were used for different tasks, e.g. for generating synthetic waveforms \cite{McSharry.2003}, waveform delineation \cite{Bock.2021}, or even human identification \cite{Israel.2005}. Evidently, this led to a plethora of proposed features and the question of which feature set is optimal for a specific task, e.g. for ECG classification \cite{mar.2011}. Regarding this application, the aim is to either assign a label to individual heart beats or to a whole recording. As an example for the latter use case, the PhysioNet/CinC Challenge 2020 posed the task to automatically assign one or multiple of $27$ classes to a large, multi-institutional database of 12-lead ECGs \cite{perez.2020}. More than $200$ teams took part with the most common algorithms being deep neural networks (DNNs).

In recent years, data-driven methods from the field of machine learning (ML) became popular, a significant percentage accounted for by DNNs \cite{Piccialli.2021}. At first many works used DNNs as classifiers and used traditional, semantic features as their input. However, recently there has been a trend towards ''end-to-end'' pipelines where the raw signal is processed and DNNs extract relevant features themselves \cite{Yang.2023, Pokaprakarn.2022, LIU2023, Le.2023, Yu.2022}. Although these methods are able to produce outstanding results and outperform conventional methods in many areas \cite{Hannun.2019, smith.2019}, a pitfall lies in the fact that they are black box models and often based on agnostic features. While they bear the theoretical potential to aid in diagnostics or treatment decisions, clinicians need to be able to comprehend their reasoning as a ''Clever Hans'' prediction \cite{clever_hans}, based on spurious or artifactual correlations, might lead to wrong decisions and adverse consequences for patients. Hence, next to issues such as inadequate performance metrics \cite{clifford.2022} and data leakage \cite{kapoor.2022}, one of the main reasons for DNNs remaining unadopted in clinical practice is missing explainability \cite{Yoon.2020, elul.2021}.

To address this need, frameworks and methods from the field of Explainable Artificial Intelligence (XAI) are developed and evaluated \cite{review_XAI}. While XAI for text and tabular input data is advancing, XAI for time series data such as biosignals is still in the need for further research \cite{Guidotti.2019}. XAI methods for DNNs include layer-wise relevance propagation (LRP) \cite{lrp}, integrated gradients (IG) \cite{integratedGrads}, and GRAD-Cam \cite{Selvaraju.2020}. However, with regard to ECG classification, these methods are usually applied qualitatively \cite{Taniguchi.2021, Bodini.2021, Sturm.2016} by showing individual recordings and corresponding XAI information, e.g. as pseudo-colored overlays. This qualitative evaluation of single recordings is rather anecdotal evidence and does not suffice the requirements for integrating DNNs in clinical practice, which needs a comprehensive characterization of models and their limitations.

\begin{figure*}[!t]
    \includegraphics[width=1\textwidth]{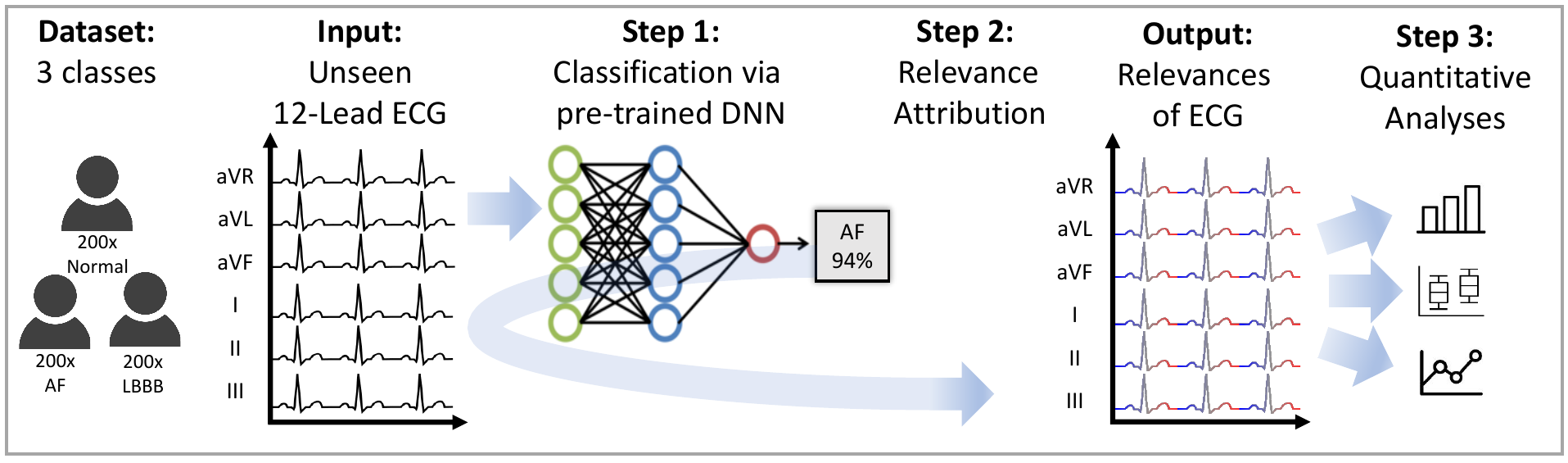}
    \caption{Overview of the processing pipeline which is applied separately to data stemming from two different databases (CPSC/PTB-XL): For each database, the data set consists of $200$ healthy controls (Normal) that are compared to patients showing AF and LBBB. Each (unseen) 12-lead ECG is fed into the pre-trained DNN and subsequently results are explored with the XAI methods, yielding a relevance score for each input sample, indicated here by blue (negative relevance score), grey (neutral), and red values (positive relevance score). We propose novel analysis methods for these scores, allowing to gain insight into the DNN's reasoning.} 
    \label{fig-overview}
\end{figure*} 

Hence, in this work, we address the unmet clinical need of missing explainability by proposing a quantitative analysis pipeline (Fig. \ref{fig-overview}) enabling an objective justification of a DNN's decision. We use a state-of-the-art, pre-trained DNN proposed by Ribeiro et al. for abnormality classification in 12-lead ECGs \cite{Ribeiro.2020} and apply attribution XAI methods to public ECG databases. In order to analyze the generalizability of this approach, we evaluate the explanatory power of different XAI methods and evaluate results on two different databases. 

The XAI methods assign to each sample of the ECG time series a relevance score reflecting how much it influenced the DNN's decision. The main contribution of this work are novel analysis methods for processing these scores. These analyses allow to gain insight into the DNN's reasoning when classifying unseen ECG signals. By mapping the results to clinical knowledge, we investigate in how far the DNN's features align with clinical knowledge. By doing so, we also propose novel visualization methods of relevance scores, allowing an intuitive and quick assessment of DNN classifications.

\section{Methods}

\subsection{Physiological Introduction}\label{medicalbg}

An ECG measures electrical activity on a patient's skin to monitor his/her cardiac cycle. It is a routine measurement in clinical settings, especially in emergency care as it allows a fast, accurate and comfortable assessment of key clinical parameters. Standard parameters derived from ECGs include heart rate, lengths between different peaks and waves, as well as the heart's electrical axis. Differences of these parameters to normal values can be interpreted as abnormalities, substantiating diagnoses. The acquisition of ECGs differs in length, e.g. $10$ s in acute care or $24$ h for Holter measurements, as well as circumstances, such as resting or exercise.

Raw ECG data is measured at equally-spaced points in time (samples) in units millivolt (mV) from multiple directions (leads) which are computed from differences in electrical potentials measured in two distinct electrodes. A standard resting ECG uses $10$ electrodes, resulting in $12$ leads, including six chest leads and six limb leads derived from electrodes on each arm and the left leg. 

The stages of the cardiac cycle, a single heart beat, are represented by characteristic waves and peaks in a P-QRS-T sequence. The P-wave represents the depolarization before the contraction of the atria which is initiated by the sinus node. The QRS-complex consists of the Q-, R-, and S-waves and corresponds to the ventricular systole, and the T-wave represents the ventricular relaxation. 

The morphology of the different waves, such as amplitude or width, as well as the intervals in between are clinically relevant. For example, atrial fibrillation (AF) is an arrhythmia based on uncoordinated electrical impulses in the atrium of the heart and a non-functioning sinus node \cite{Hindricks.2021} that can be diagnosed from ECGs. Criteria for diagnosis are absence of P-waves, as they are initiated by the sinus node, and irregular RR intervals \cite{Hindricks.2021}. However, repeating fibrillatory waves (f-waves) mimic P-waves and can usually be observed best in leads V1-6, especially V1 \cite{Bollmann.2006}. Another abnormality is left bundle branch block (LBBB), where the cardiac conduction through the left bundle branch is compromised downstream from lesions of the His bundle or its derivatives. LBBB criteria for ECGs include unusually wide QRS-complexes with the ST-segment and T-waves pointing in opposite direction \cite{Tan.2020}. I, aVL, V5 and V6 are left-sided leads, where broad notched or slurred R-waves can be observed, while Q waves are absent \cite{Tan.2020}. Both, AF and LBBB, can be diagnosed by ECG acquisition with a reduced number of leads, but the gold standard for diagnosis is 12-lead ECG \cite{harris.2012}.

\subsection{Technical Background}
Ribeiro et al. published a residual network (ResNet) trained on more than two million ECGs from a Brazilian telehealth network, showing F1-scores of more than $80$ \% for classification of six ECG abnormalities. The output from convolutional layers in each of four residual blocks are fed into a fully connected layer with sigmoid activation function, yielding independent probabilities for six classes of ECG abnormalities \cite{Ribeiro.2020c}. Thresholds calculated for the final classifications are available on GitHub\footnote{\url{https://github.com/antonior92/automatic-ecg-diagnosis/blob/master/generate_figures_and_tables.py}, commit 89f929d, line 121}. In previous work, we demonstrated methods and results reproducibility with local data \cite{Bender.2021}.

The model accepts a matrix with dimensions $N \times 4096 \times 12$ with $4096$ and $12$ defining the number of samples and leads, respectively. $N$ denotes the number of recordings to be processed. The model outputs a matrix with dimensions $N\times6$ assigning probabilities for six ECG abnormalities, namely first degree AV block, right bundle branch block, LBBB, sinus bradycardia, AF and sinus tachycardia.

In medical applications such as ECG diagnostics it is important for clinicians to understand the reasoning of a DNN. XAI methods build a wrapper around the black box model, giving insight into possible features that led to the DNN's output. In this paper, we focus on two state-of-the-art attribution methods, IG and LRP.

\subsubsection{Integrated Gradients}
IG attribute the prediction of a neural network on unseen data to its input features. However, IG use a baseline input for attribution calculation. The authors \cite{integratedGrads} motivate this by noting that if we assign blame to something, we implicitly consider the absence of it as a baseline for comparing outcomes.

IG are calculated as follows: Let $f$ be a function that represents a neural network, $x$ the input at hand, and $\Tilde{x}$ the baseline input. The IG are defined as the path integral of the gradients along the straight-line path from the baseline $\Tilde{x}$ and input $x$. The straight-line path can easily be written down as $\Tilde{x}+\alpha(x-\Tilde{x})$ for $\alpha\in[0,1]$. The integrated gradient for the $i$-th input dimension is defined as

\begin{equation}\label{integratedGrad}
    \operatorname{IG}_i(x):=(x_i-\Tilde{x_i})\cdot\int_0^1\frac{\partial f(\Tilde{x}+\alpha(x-\Tilde{x}))}{\partial x_i}d\alpha,
\end{equation}

where $\frac{\partial f(x)}{\partial x_i}$ is the gradient of $f(x)$ along the $i$-th dimension.

The property of the LRP methods that the relevance scores of the input can be summed up and approximate the prediction score (see (\ref{summationToPredictionScore})) can also be proven for IG by using the fundamental theorem of calculus for path integrals. This states that if $f:\mathbb{R}^n\rightarrow\mathbb{R}$ is differentiable almost everywhere\footnote{This means $f$ is continuous everywhere and the partial derivative of $f$ along each input dimension is Lebesgue integrable. This holds for most neural networks using Sigmoid, ReLU, or Pooling functions.} then

\begin{equation}\label{sumOverIntGrads}
    \sum_{i=1}^n \operatorname{IG}_i(x)=f(x)-f(\Tilde{x}).
\end{equation}

For a baseline $\Tilde{x}$ with prediction $f(\Tilde{x})$ near zero, we can see that the sum over the IG in (\ref{sumOverIntGrads}) also approximates the prediction score $f(x)$ similar to how the sum over the relevance scores calculated by LRP approximates the prediction score $f(x)$ in (\ref{summationToPredictionScore}). This property is termed \textit{completeness} in \cite{integratedGrads}.

For computing IG the integration is replaced by a sum over sufficiently small intervals along the straight-line path

\begin{equation}\label{approxIntGrads}
    \operatorname{IG}_i^{approx}(x):=(x_i-\Tilde{x_i})\cdot\sum_{k=1}^m\frac{\partial f(\Tilde{x}+\frac{k}{m}(x-\Tilde{x}))}{\partial x_i}\frac{1}{m}.
\end{equation}

\subsubsection{Layer-wise Relevance Propagation}
LRP tries to explain the output $f(x)$ made by a classifier $f$ with respect to an input $x$ by decomposing the output $f(x)$ in such a way that

\begin{equation}\label{summationToPredictionScore}
f(x)\approx\sum_{d=1}^V R_d,    
\end{equation}

where $V$ is the input dimension. $R_d>0$ would then indicate the presence of the structure which is to be classified and $R_d<0$ would indicate its absence.

Propagation of relevance scores works as follows: Let $R_j^{(\ell+1)}$ be a known relevance score of a certain neuron $j$ in the $\ell+1$-th layer of a neural network, for a classification decision $f(x)$. The decomposition of the relevance score $R_j^{(\ell+1)}$ in terms of messages $R_{i\leftarrow j}$ sent to neurons of the previous layer $\ell$ must hold the conservation property

\begin{equation}\label{conservationProp}
    \sum_i R_{i\leftarrow j}^{(\ell,\ell+1)} = R_{j}^{(l+1)},
\end{equation}

where $\sum_i$ describes the sum over all neurons in the $\ell$-th layer of the neural network.\\ 
One possible relevance decomposition that satisfies (\ref{conservationProp}) would be to use the ratio of local and global activations:

\begin{align}\label{basicLRP}
    R_{i\leftarrow j}^{(\ell,\ell+1)} =& \frac{x_i^{(\ell)}\omega_{ij}^{(\ell,\ell+1)}}{\sum_k x_k^{(\ell)}\omega_{kj}^{(\ell,\ell+1)}+b_j^{(\ell)}}R_j^{(\ell+1)}\nonumber \\ =& \frac{x_i^{(\ell)}\omega_{ij}^{(\ell,\ell+1)}}{z_j}R_j^{(\ell+1)},
\end{align}

where $x_i$ is the activation (calculated by a non-linear activation function) of the $i$-th neuron in the $\ell$-th layer, $w_{ij}^{(\ell,\ell+1)}$ is the weight connecting neuron $i$ in the $\ell$-th layer to neuron $j$ in the $\ell+1$-th layer, $b_j^{(\ell)}$ is a bias term, and $\sum_k$ describes the sum over all neurons in the $\ell$-th layer.

A problem with (\ref{basicLRP}) is that if $z_j$ gets very small, the relevance scores $R_{i\leftarrow j}$ can get infinitely large. To overcome this problem, the authors of \cite{lrp} introduced a stabilizer $\epsilon\geq0$:

\begin{equation}\label{lrpEpsilon}
    R_{i\leftarrow j}^{(\ell,\ell+1)} = \begin{cases}
    \frac{x_i^{(\ell)}\omega_{ij}^{(\ell,\ell+1)}}{z_j + \epsilon}R_j^{(\ell+1)}, & \text{if $z_j\geq0$},\\
    \frac{x_i^{(\ell)}\omega_{ij}^{(\ell,\ell+1)}}{z_j - \epsilon}R_j^{(\ell+1)}, & \text{if $z_j<0$}.
  \end{cases}
\end{equation}

As we can see in (\ref{lrpEpsilon}), if $\epsilon$ becomes very large, the relevance scores will tend to zero which poses another problem. To counteract this, a different treatment of positive and negative activations $x_i$ is proposed in \cite{lrp}. Let $z_j^+$ and $z_j^-$ denote the positive and negative part of $z_j$ such that $z_j^+ +z_j^-=z_j$. The same notation will be used for the positive and negative parts of $x_i^{(\ell)}\omega_{ij}^{(\ell,\ell+1)}$. Relevance decomposition can now be defined by

\begin{align}\label{lrpAlphaBeta}
    R_{i\leftarrow j}^{(\ell,\ell+1)} =R_j^{(\ell+1)}\cdot&\left(\alpha\cdot \frac{\left(x_i^{(\ell)}\omega_{ij}^{(\ell,\ell+1)}\right)^+}{z_j^+}\right.\nonumber \\
    &\hspace{0.3cm}\left.+\beta\cdot \frac{\left(x_i^{(\ell)}\omega_{ij}^{(\ell,\ell+1)}\right)^-}{z_j^-}\right),
\end{align}

where $\alpha+\beta=1$.\\
A different propagation rule has been proposed by \cite{Montavon_2017} for real valued inputs that redistributes relevance scores according to the square magnitude of the weights:

\begin{equation}\label{lrpOmega2}
    R_{i\leftarrow j}^{(\ell,\ell+1)} = \frac{\left(\omega_{ij}^{(\ell,\ell+1)}\right)^2 }{\sum_k \left(\omega_{kj}^{(\ell,\ell+1)}\right)^2}R_j^{(\ell+1)}
\end{equation}

Other papers such as \cite{Samek2017UnderstandingAC} and \cite{Kohlbrenner} propose a combination of different decomposition rules for different layer types, like (\ref{lrpEpsilon}) for fully connected layers to truthfully represent the decisions made via the layers' linear mapping and (\ref{lrpAlphaBeta}) for convolutional layers with ReLU activation functions to separately handle the positive and negative parts of $x_i^{(\ell)}\omega_{ij}^{(\ell,\ell+1)}$.

\subsection{Experimental Design}
Fig. \ref{fig-overview} shows an overview of our DNN and XAI pipeline applied in this work. This pipeline is run separately on data stemming from two different databases.

\subsubsection{Databases}
The data set for our main analysis stems from the CPSC2018 database\footnote{\url{https://storage.cloud.google.com/physionet-challenge-2020-12-lead-ecg-public/PhysioNetChallenge2020_Training_CPSC.tar.gz}} acquired in eleven Chinese hospitals containing 12-lead ECGs with a ground truth provided by human experts \cite{Liu.2018}. Additionally, we validate the generalizability of our results using the PTB-XL database \cite{Wagner.2020} as described in sec. \ref{method-databases}. An overview of the properties of both databases is shown in Tbl. \ref{tab:datasets}.

For our main analysis on the CPSC database, we use a subset of $200$ each for AF, LBBB and healthy subjects showing normal signals, resulting in $N=600$ recordings. We investigate these two classes as AF is defined by an abnormal heart rhythm, i.e. irregular distances between heart beats, and therefore it can only be diagnosed by analyzing multiple heart beats. In contrast, LBBB can be diagnosed by a single heart beat as it is characterized by distinct morphological features, e.g. a notched QRS-complex. 

\begin{table}[]
\caption{Properties of CPSC \cite{Liu.2018} and PTB-XL \cite{Wagner.2020} }
\label{tab:datasets}
\begin{tabular}{l||c|c|c||c|c|}
          & \# ECGs & Duration & Sampling& \# Patients & Country\\
\hline
CPSC & $9,831$   & $6-60$s & $500$~Hz & $9,458$ & PRC \\
\hline
PTB-XL    & $21,799$  & $10$s & $500$~Hz & $18,869$ & GER \\    
\end{tabular}
\end{table}

\subsubsection{Processing pipeline}
All recordings were resampled to $400$ Hz and trimmed or zero-padded to $4096$ samples. In the remainder of this work, we denote a single ECG sample as $E_{n,j,k}$ with $n = \{0,1,\dots 599\}$ representing the recording index, $j = \{0,1,\dots 4095\}$ representing samples, and $k= \{0,1,\dots 11\}$ representing leads.
Regarding data processing\footnote{All computations are implemented using Python v3.6.8 and the libraries iNNvestigate v1.0.9, Tensorflow v1.12.0, neurokit2 V0.1.7, and h5py v2.10.0.}, each ECG signal is fed to the model by Ribeiro et al. \cite{Ribeiro.2020b} for classification, resulting in 
a matrix with dimensions $N\times6$ assigning probabilities for six ECG abnormalities. In the following, we define $\{C_n \in \mathbb{R} ~\vert~ 0 \le C_n \le 1 \}$ indicating the prediction score of the model with sigmoid activation, representing the classification probability. We utilize the package iNNvestigate \cite{Alber.2019}, which implements multiple XAI methods, to compute relevance scores for each sample of the input ECGs. 
We use the XAI methods IG and LRP with the IG implementation being with baseline input zero and interval size $m=64$, after changing the activation of the DNN's last layer to linear. Sigmoid activation does not change the ranking order of the predicted classes, but might obfuscate the true confidence of the model's individual class predictions\footnote{\url{https://github.com/albermax/innvestigate/issues/84}, accessed: October 14, 2022}.

The XAI methods assign a relevance score $R_{j,k} \in \mathbb{R}$ to each input sample of a classified ECG recording. By computing this for all $N$ recordings we obtain $R_{n,j,k}$ with the same dimensions as our input ECG data $E_{n,j,k}$. Both are the basis for our analysis to compare features embedded in the DNN model to clinically-relevant criteria.
We analyze the obtained relevance scores $R_{n,j,k}$ with three novel quantitative methods and one qualitative method as described in the following sections. 
With each new analysis, we take more details into account. While in the first analysis relevance scores are binned to each class, in the second analysis we split relevance scores w.r.t. their lead and in the third analysis w.r.t. lead and heart beats.

\subsubsection{Binned and Average Relevance Scores Over Class}
We first analyze relevance scores for all $200$ normal, $200$ LBBB, and $200$ AF recordings separately and bin the values for their respective class, allowing us to compare the overall distribution of $R_{n,j,k}$ for the different classes.

We then aggregate all leads of each recording $n$ into

\begin{equation}\label{meanSignal}
    M_{n} \vcentcolon= \frac{1}{J~K}\sum\limits_{k=0}^{K-1} 
    \sum\limits_{j=0}^{J-1} R_{n,j,k}
    ,
\end{equation}

with $K=12$ and $J=4096$. $R_{n,j,k}$ takes positive or negative values, hence a higher $M_{n}$ is associated with a higher prediction score, termed \textit{completeness} in \cite{integratedGrads}. Here, the prediction score is the output of the model with linear activation.

\subsubsection{Average Relevance Scores Over Class and Lead}

We aggregate relevance scores for each lead $k$ and recording $n$ in
\begin{equation}\label{meanLead}
    M_{n,k} \vcentcolon= \frac{1}{J}\sum\limits_{j=0}^{J-1} R_{n,j,k},
\end{equation}
with $J=4096$. This allows for comparing the distribution of $R_{n,j,k}$ w.r.t. class and ECG leads and thus the importance of the individual ECG leads for the DNN. This is required as the different leads show different morphologies and signal shapes that might cancel out in the first analysis.

\subsubsection{Average Relevance Scores Over Class, Lead, and Beats}

In the first two analysis methods, time information is lost. However, for explaining the DNNs decision this is crucial as we need to compare whether the agnostic features trained by the DNN reflect the clinical features described in section \ref{medicalbg} such as missing P-waves, unusually wide QRS-complexes etc. 

Analyzing individual ECG records gives only anecdotal evidence. Therefore, we perform a two-step averaging procedure which averages the information over several recordings while preserving time information. First, for each ECG record and lead, we use the concept of ''average beats'' \cite{hamilton.1991} by splitting the whole signal into individual heart beats with the \textit{ecg\_segment()} function of neurokit2. We average them into a single, time-aligned representative beat for each lead. Then we use the exact same indices of the heart beats and perform the same steps on the relevance scores $R_{n,j,k}$, yielding an ''average relevance score''. All average beats and average relevance scores are then averaged for a given class. All segments are of equal size for one recording, hence we fill segments overlapping start or end of the recording with zeros. Finally, amplitudes are normalized to $[-1,1]$. For scatter plot visualizations, relevance scores are upsampled by a factor of $5$.

\subsubsection{Qualitative Analysis of XAI Relevance Scores}

The results of all processed ECG signals were visualized as heatmap-colored scatter plots for each lead, after a normalization of the output to $[-1,1]$, keeping the center of the values at zero. Furthermore, these relevance score plots were evaluated by an experienced cardiologist.

\subsubsection{Comparison Between Databases}
\label{method-databases}
To evaluate the the generalizability of our processing pipeline, we evaluate results on another publicly-available dataset. For this task we use PTB-XL \cite{Wagner.2020} which is which is an older public database acquired between October 1989 and June 1996 in Germany. Therefore, the ECG measurement equipment and subject's origin are completely different to the CPSC database and additionally there is the chance of different clinical guidelines being in practice for the annotation by cardiologists.

\subsubsection{Comparison Between XAI Methods}\label{method-comparison}

Since both methods, IG and LRP, differ substantially in their approach on how to calculate relevance scores for the input, we believe that using both methods will help uncover important information about why the DNN made certain decisions. Hence, we compare IG results to LRP using the following LRP decomposition rules implemented in the iNNvestigate \cite{Alber.2019} package:
\begin{itemize}
    \item[a)] The $\epsilon$-LRP decomposition (see (\ref{lrpEpsilon})) with $\epsilon=1e-07$.
    \item[b)] The $\alpha\beta$-LRP decomposition (see (\ref{lrpAlphaBeta})) with $\alpha=1$ and $\beta=0$.
    \item[c)] The $\omega^2$-LRP decomposition (see \ref{lrpOmega2})).
    \item[d)] The combination of $\alpha\beta$-LRP decomposition (see (\ref{lrpAlphaBeta})) with $\alpha=1$ and $\beta=0$ for convolutional layers and $\epsilon$-LRP decomposition (see (\ref{lrpEpsilon})) with $\epsilon=0.1$ for fully connected layers.
\end{itemize}

The sigmoid function (used in the output layer) maps from $\mathbb{R}$ to $\mathbb{R}^+$ and thus inverts the signs of all negative values, as well as scales all values into the interval of $[0,1]$. This results in only small and positive values being backpropagated by the LRP method possibly resulting in small and only positive relevance scores. Thus we compared these relevance scores to those obtained by using a linear output in the last layer. 
Since both activations yield similar results when compared visually in heatmaps, we decided to continue with linear activation, to avoid the possible sign flip.

\subsection{Ethics approval}
Human subject research: This work only makes use of public data and does not contain any additional information involving human participants obtained by the authors.

\section{Results}

After processing recordings with the DNN, $C_n$ is the probability that a recording $n$ shows the interrogated abnormality. The recording is classified as this abnormality if $C_n$ is higher than a threshold defined by Ribeiro et al., which is $0.39$ for AF and $0.05$ for LBBB. Applying an XAI method results in a relevance score $R_{n,j,k} \in \mathbb{R}$ for each input sample of a classified ECG with $j = \{0,1,\dots 4095\}$ representing sample index, and $k= \{0,1,\dots 11\}$ representing the lead.

\subsection{Average Relevance Scores Over Class}

\begin{figure*}[!t]
    \centering
    \begin{subfigure}[b]{0.48\textwidth}
        \centering
        \includegraphics[width=1\textwidth]{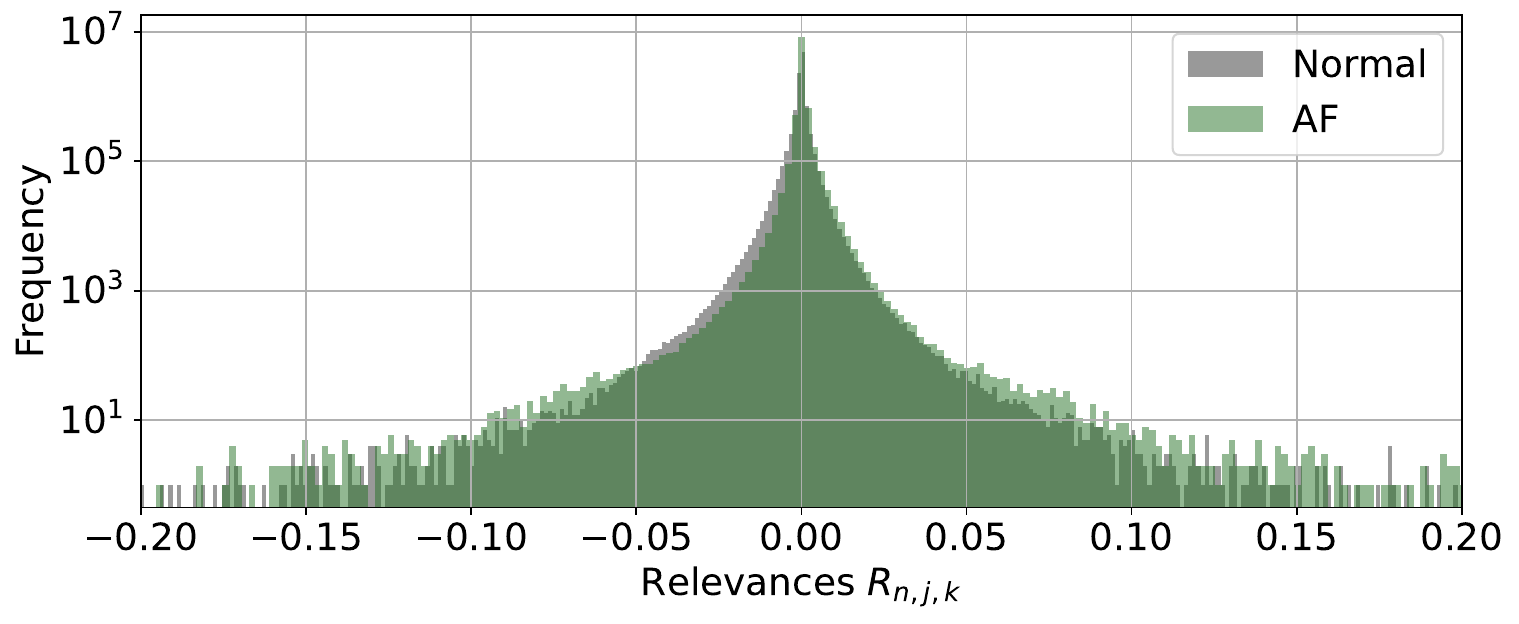}
    	\caption{Normal and AF recordings. Colors denote ground truth label of data set. Values for AF range from $[-0.5,0.5]$ and values for normal recordings from $[-0.3,0.4]$.}
    	\label{fig-hist-AFvsSR}
    \end{subfigure}
    \hfill
    \begin{subfigure}[b]{0.48\textwidth}
        \centering
    	\includegraphics[width=1\textwidth]{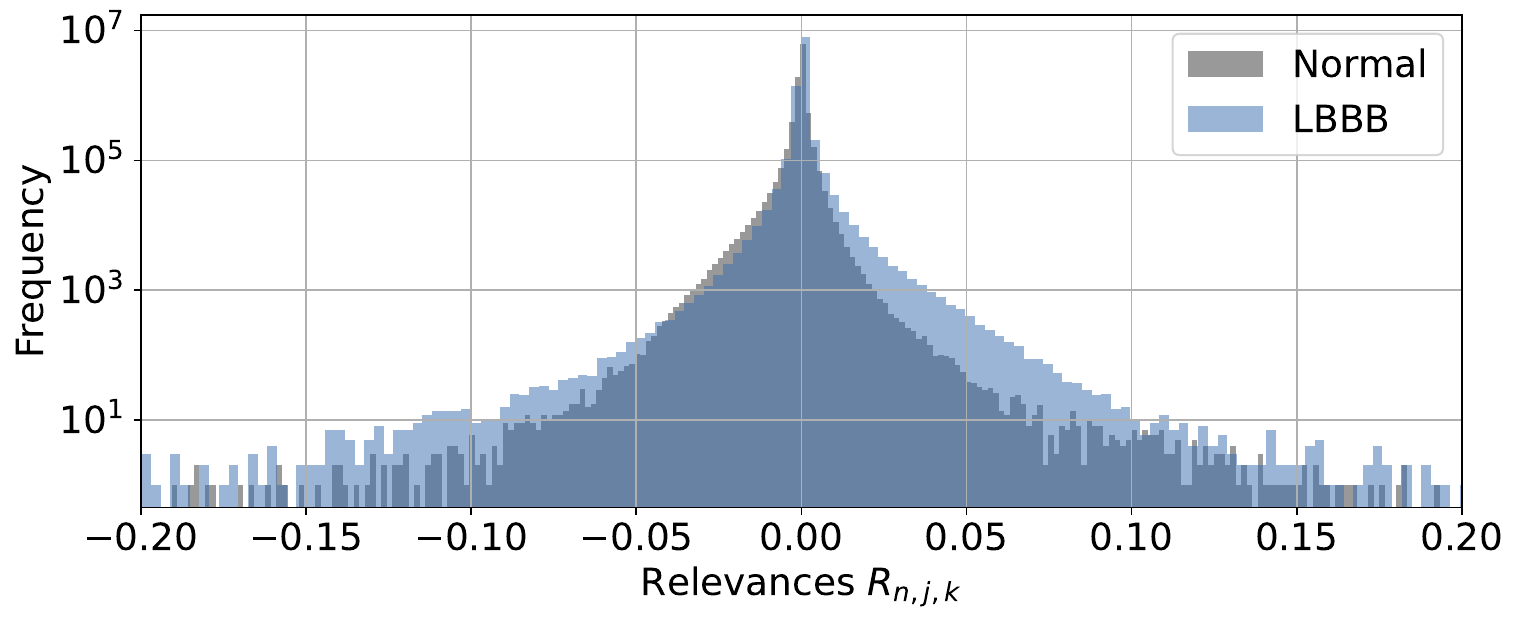}
    	\caption{Normal and LBBB recordings. Colors denote ground truth label of data set. Values for LBBB range from $[-0.6,0.9]$ and values for normal recordings from $[-0.4,0.5]$.}
    	\label{fig-hist-LBBBvsSR}
    \end{subfigure}
	\caption{Distribution of IG relevance scores $R_{n,j,k}$.  To increase visibility, x-axes are limited to $[-0.20,0.20]$.}
	\label{fig-hist}
\end{figure*}

\begin{figure*}[!t]
    \centering
    \begin{subfigure}[b]{0.75\textwidth}
        \centering
        \includegraphics[width=1\textwidth]{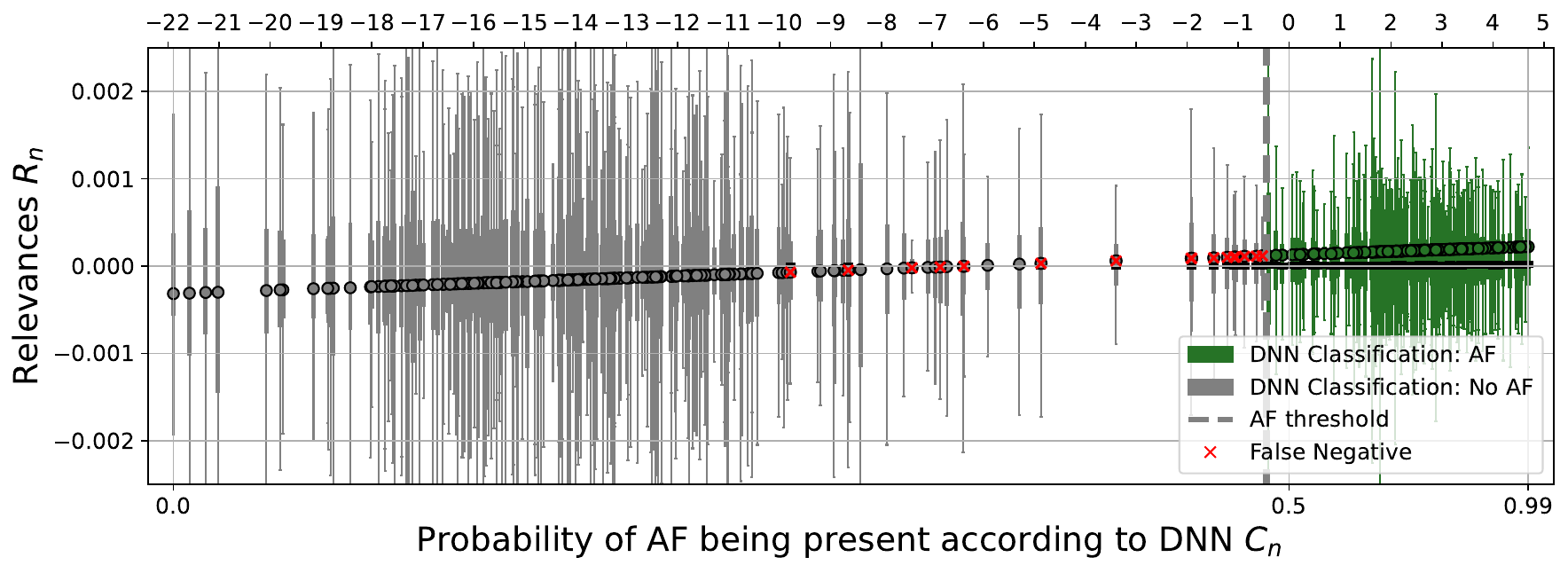}
    	\caption{Atrial Fibrillation}
    	\label{fig-boxplotR-AFvsSR}
    \end{subfigure}
    \begin{subfigure}[b]{0.75\textwidth}
        \centering
    	\includegraphics[width=1\textwidth]{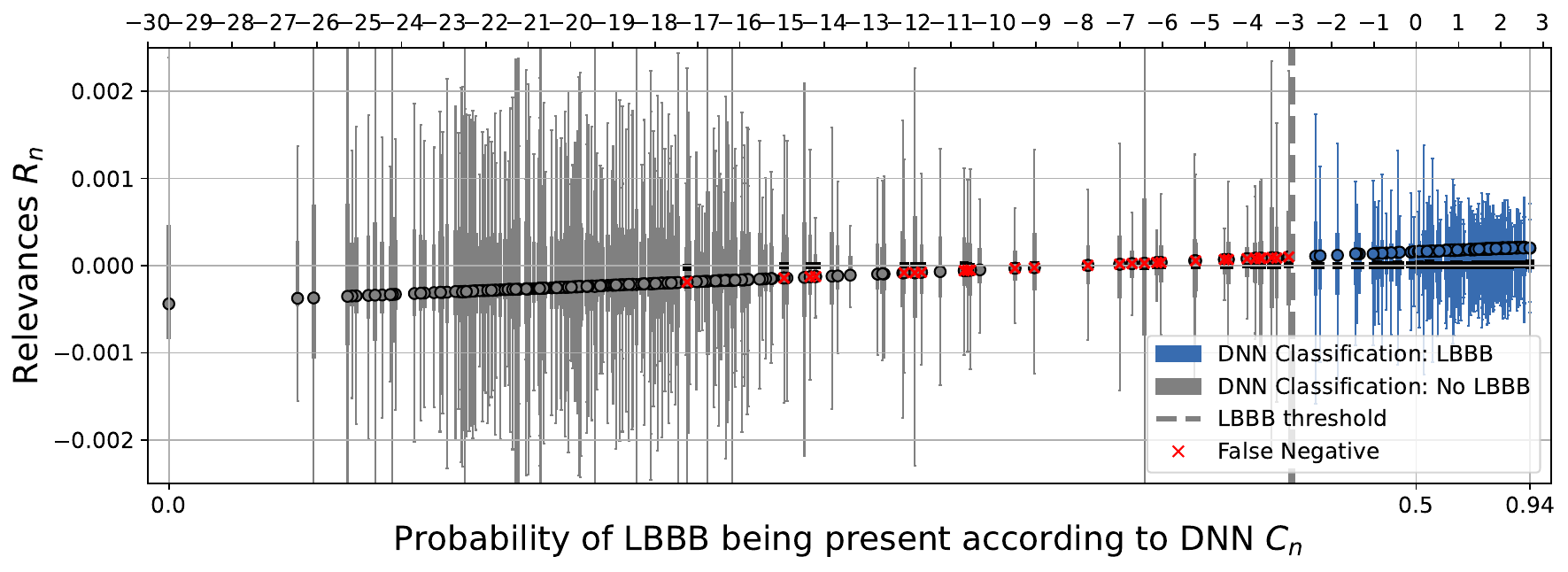}
    	\caption{Left Bundle Branch Block}
    	\label{fig-boxplotR-LBBBvsSR}
    \end{subfigure}
	\caption{Distribution of $R_{n}$ computed with IG for each recording as single boxplot. The bottom x-axis represents sigmoid activation output of the DNN, while the upper x-axis represents the output with linear activation. Boxplot colors denote DNN classification results and red crosses indicate false negatives.}
	\label{fig-boxplotR}
\end{figure*}

\begin{figure*}[!t]
    \centering
    \begin{subfigure}[b]{0.48\textwidth}
        \centering
        \includegraphics[width=1\textwidth]{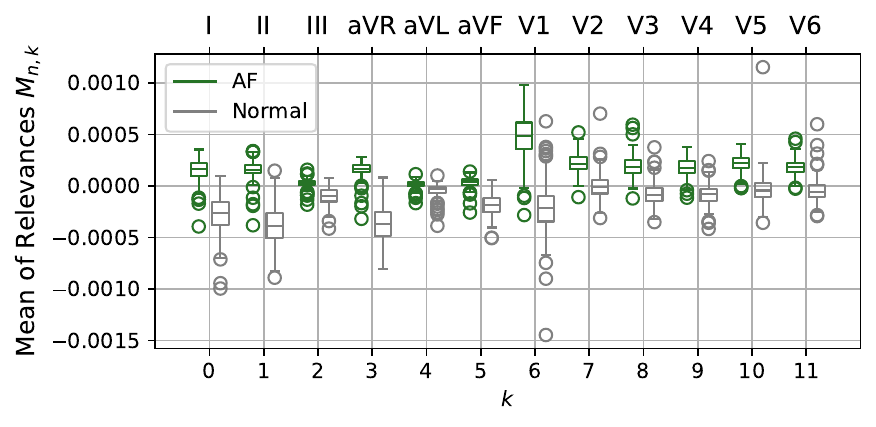}
    	\caption{AF classification}
    	\label{fig-boxplot-AFvsSR}
    \end{subfigure}
    \hfill
    \begin{subfigure}[b]{0.48\textwidth}
        \centering
    	\includegraphics[width=1\textwidth]{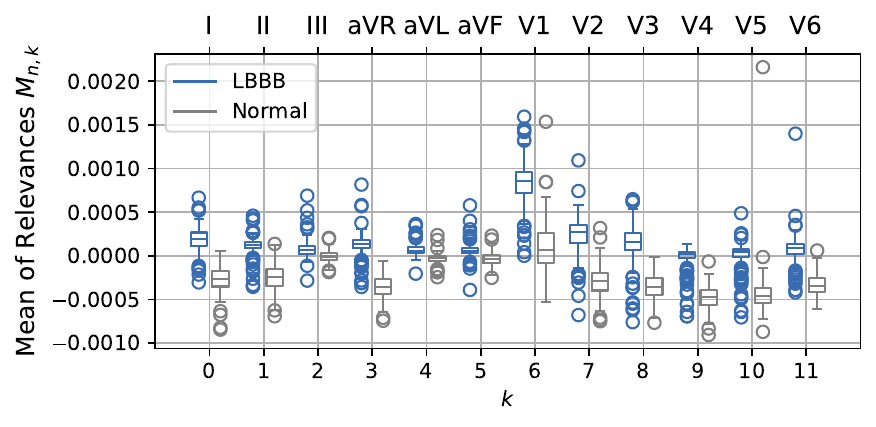}
    	\caption{LBBB classification}
    	\label{fig-boxplot-LBBBvsSR}
    \end{subfigure}
    	\caption{Distribution of $M_{n,k}$ computed with IG w.r.t. ECG leads, colors denoting ground truth label. For AF classification (a) and LBBB classification (b) boxplots show that the abnormal mean is higher for each lead with the highest difference in V1.}
    	\label{fig-boxplot}
\end{figure*}

The mean of the distributions of IG relevance scores $R_{n,j,k}$ for each class (Fig. \ref{fig-hist}) is close to zero, representing that the majority of ECG samples is not relevant for the DNN's decision. Distributions for both abnormalities are almost similar to normal recordings, although they are slightly broader and shifted to positive values. For LBBB, in the range $[0.0,0.10]$ there is a large number of more positive relevance scores compared to normal recordings (Fig. \ref{fig-hist-LBBBvsSR}).

The relevance scores of individual recordings are again centered close to zero and rather equally-distributed (Fig. \ref{fig-boxplotR}). In general, AF shows larger values in positive and negative direction compared to LBBB. While the median value is always very close to zero, the mean value of relevance scores is increasing with increasing $C_n$. For AF classification (Fig. \ref{fig-boxplotR-AFvsSR}) a large amount of normal recordings correctly classified as not showing AF have a $C_n$ near $0$ and correctly classified AF recordings are near $1$. In between is a ''transition area'' with nine false negative classifications in $[0.1, 0.39[$. The remaining seven false negatives show $M_{n}$ values close to zero. LBBB has similar properties to AF, although there is no visible transition area and the values are not as close to $1$ (Fig. \ref{fig-boxplotR-LBBBvsSR}).

\subsection{Average Relevance Scores Over Class and Lead}

Analyzing model results of each lead $k$ for AF classification (Fig. \ref{fig-boxplot-AFvsSR}), mean relevance scores showed medians of $0.0002$,$-0.0001$ and ranges of $[-0.0002,0.0010]$ and $[-0.0014,0.0012]$ for AF and normal recordings, respectively. For LBBB classification (Fig. \ref{fig-boxplot-LBBBvsSR}), medians were $0.0001$,$-0.0002$ and ranges were $[-0.0008,$ $0.0016]$ and $[-0.0009,0.0022]$ for LBBB and normal recordings, respectively. For each lead, the mean relevance scores were significantly higher for both abnormalities compared to normal recordings, with a Wilcoxon-Rank-Sum-Test and p-value $<$ 0.01. Particularly, lead V1 shows the highest difference in median values.

\begin{figure*}[t!]
    \centering
	\includegraphics[width=0.95\textwidth]{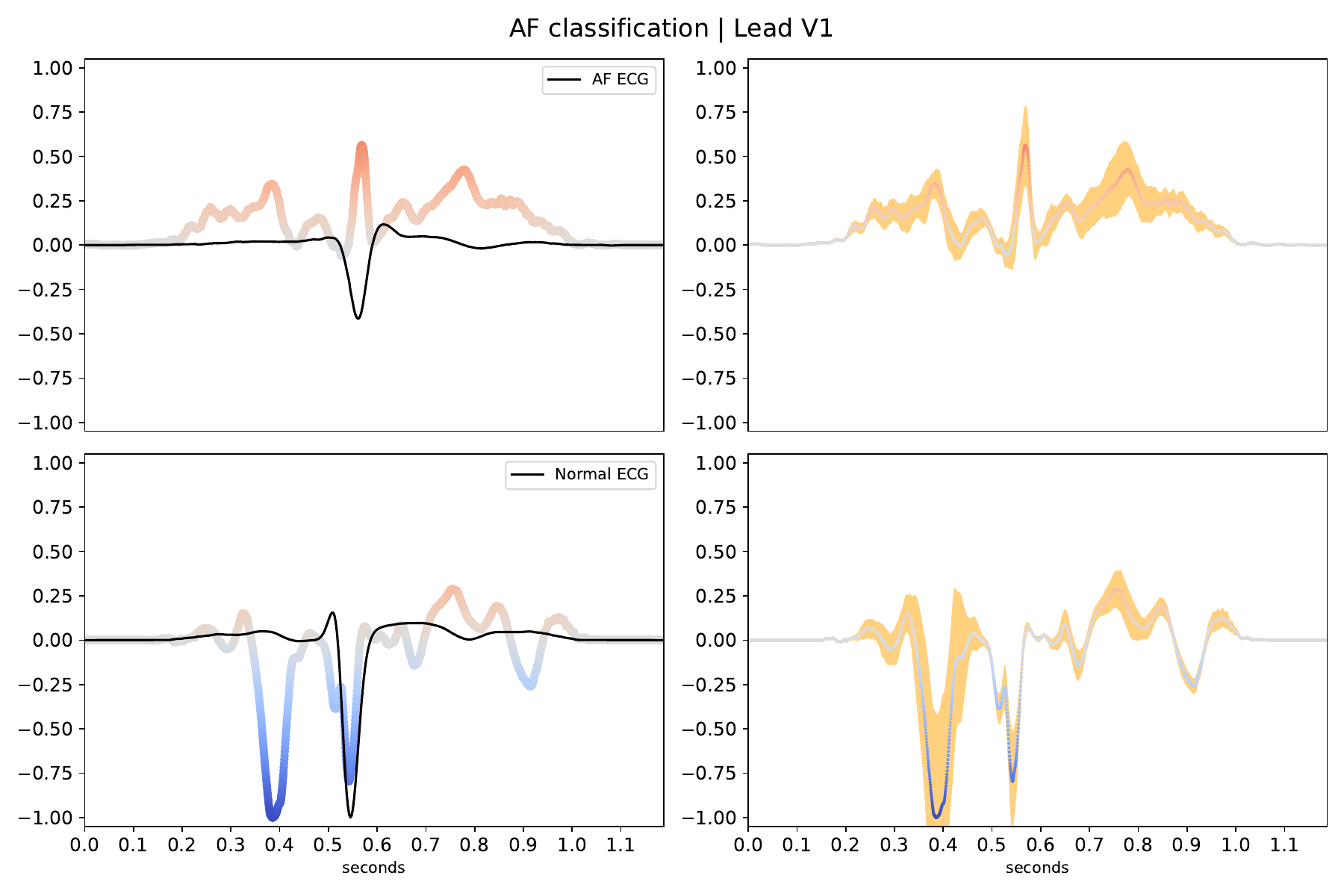}
	\caption{Left column: Average beats (black curves) and IG relevance scores for lead V1 in AF classification. Abnormal ECGs show positive relevance scores (red) distributed over the whole P-QRS-T-cycle, negative relevance scores (blue) on normal recordings cover QRS-complexes and especially P-waves. Right column: Instead of average beats, the variance of relevance scores across recordings is shown (orange).}
	\label{fig-avg-AFvsSR}
\end{figure*}

\subsection{Average Relevance Scores Over Class, Lead, and Beats}

Average beats over $200$ recordings show mostly positive relevance scores for both abnormalities, and mostly negative relevance scores for normal recordings for both classifications (Fig. \ref{fig-avg-AFvsSR}). 

\begin{figure*}[t!]
    \centering
	\includegraphics[width=0.95\textwidth]{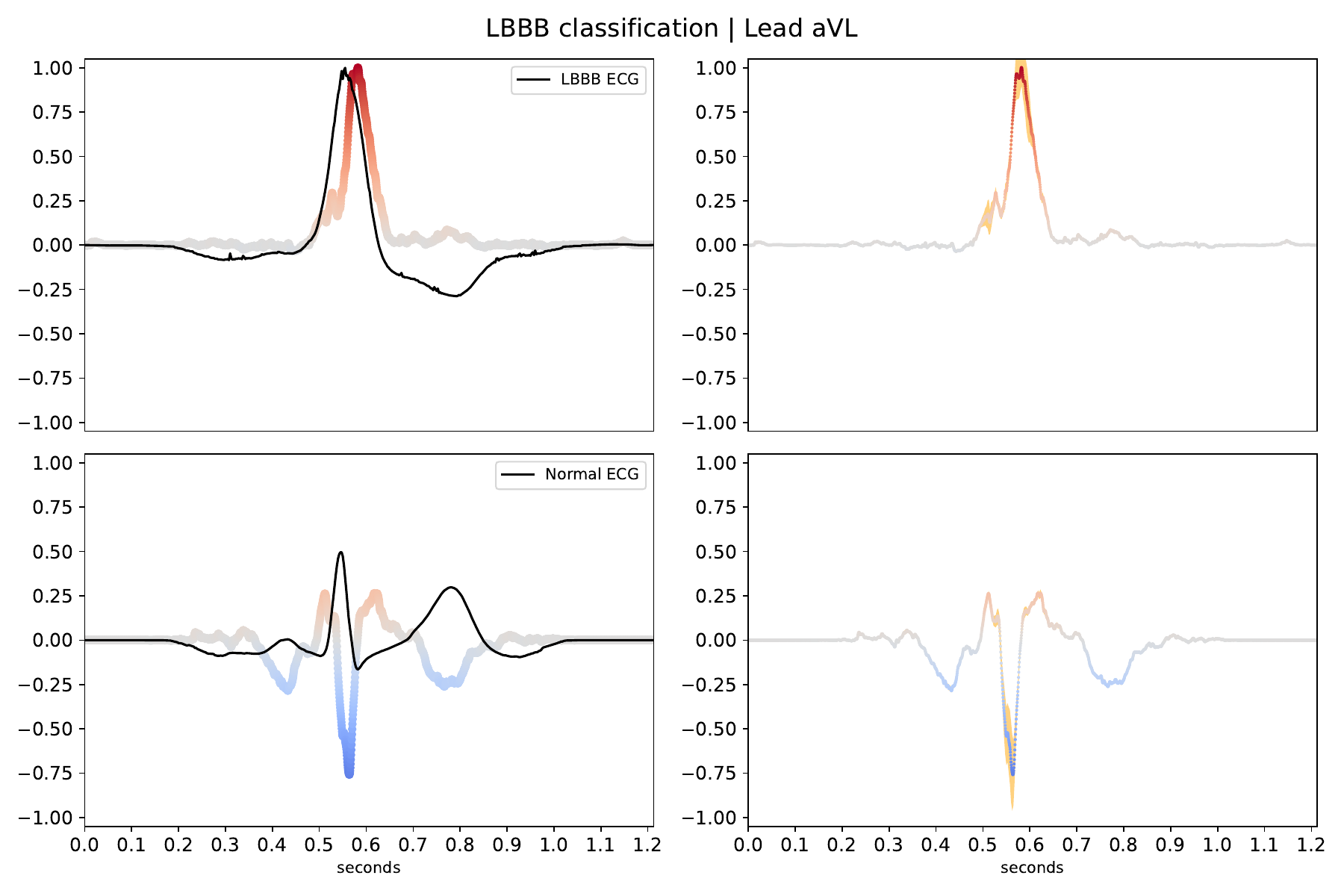}
	\caption{Left column: Average beats and IG relevance scores for lead aVL in LBBB classification. Abnormal ECGs show positive relevance scores (red) on QRS-complexes; negative scores (blue) on normal recordings can be seen on P- and T-waves. Right column: Instead of average beats, the variance of relevance scores across recordings is shown (orange).}
	\label{fig-avg-LBBBvsSR}
\end{figure*}

When classifying AF, QRS-complexes are the most relevant areas, especially R-peaks. For normal recordings, we observed high negative values for the area of P-waves as well. Negative values of normal recordings are higher compared to positive values of AF recordings. For LBBB classification, QRS-complexes are most relevant as well (Fig. \ref{fig-avg-LBBBvsSR}). Furthermore, the concentration of high absolute relevance scores on specific waves or peaks is clearer, such as the negative T-wave in LBBB, assigned with negative relevance scores when positive in normal recordings. In contrast, for AF many smaller relevance scores with higher variance are distributed on the whole beat.

\subsection{Qualitative Analysis}

\begin{figure}[t!]
    \includegraphics[width=1\columnwidth]{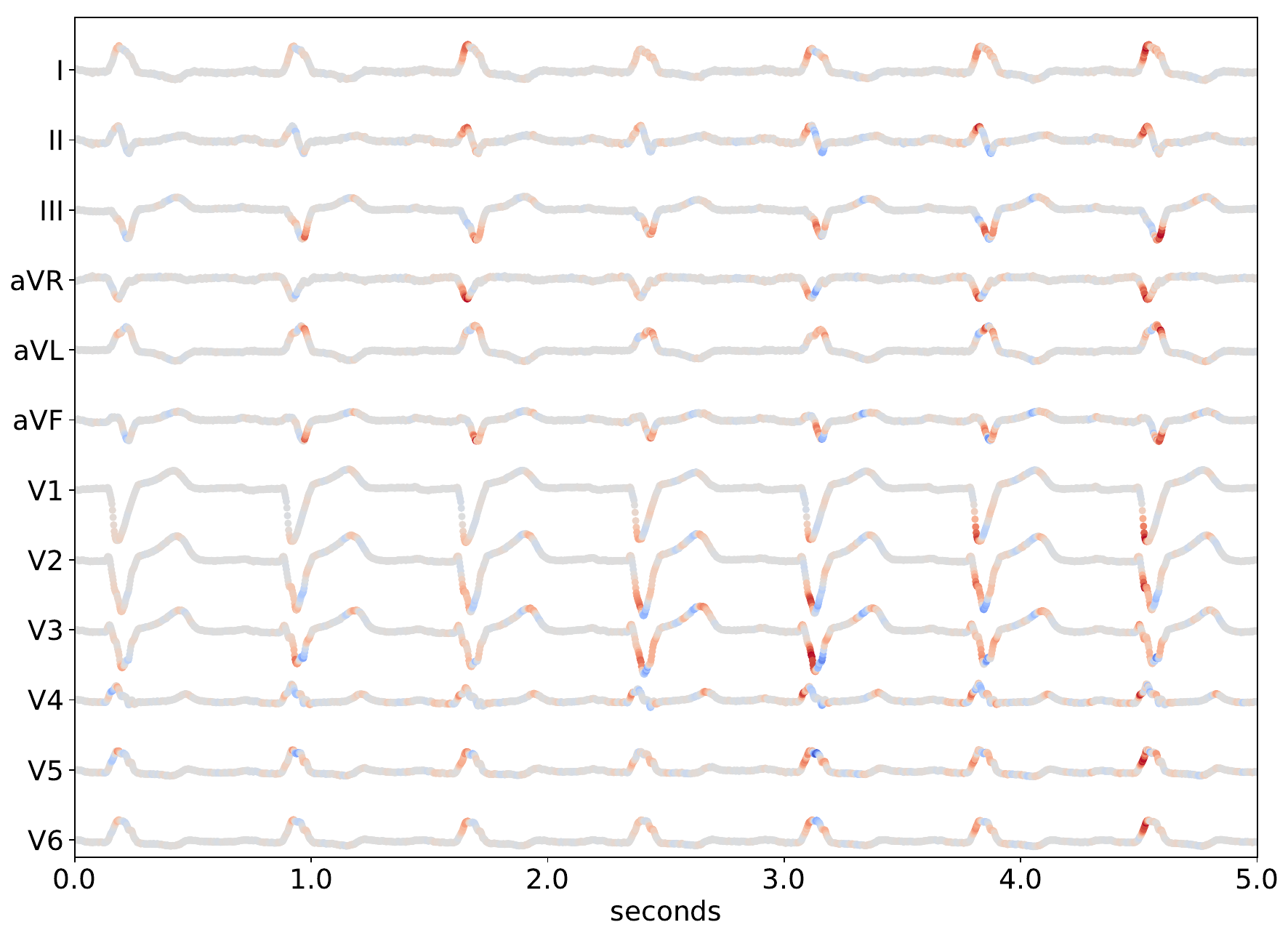}
    \caption{Positive (red) and negative (blue) relevance scores calculated with IG on correctly classified electrocardiogram (LBBB: $\sim0.871$) from CPSC data set (ID A0977). Relevance scores normed to $[-1,1]$ per lead. \label{fig-relevances-86}}
\end{figure}

We observed clusters of high absolute relevance scores in the area of QRS-complexes during visual inspection of single recordings visualized as heatmap (Fig. \ref{fig-relevances-86}). For LBBB, IG seems to focus on negative S-waves and prolonged ST-segments in lead V1. Occasionally, broad and notched R-waves were also marked relevant. On the contrary, for AF recordings, the relevant parts were usually R-waves and in rare instances areas with missing P-waves.

\begin{figure}[t!]
    \includegraphics[width=\columnwidth]{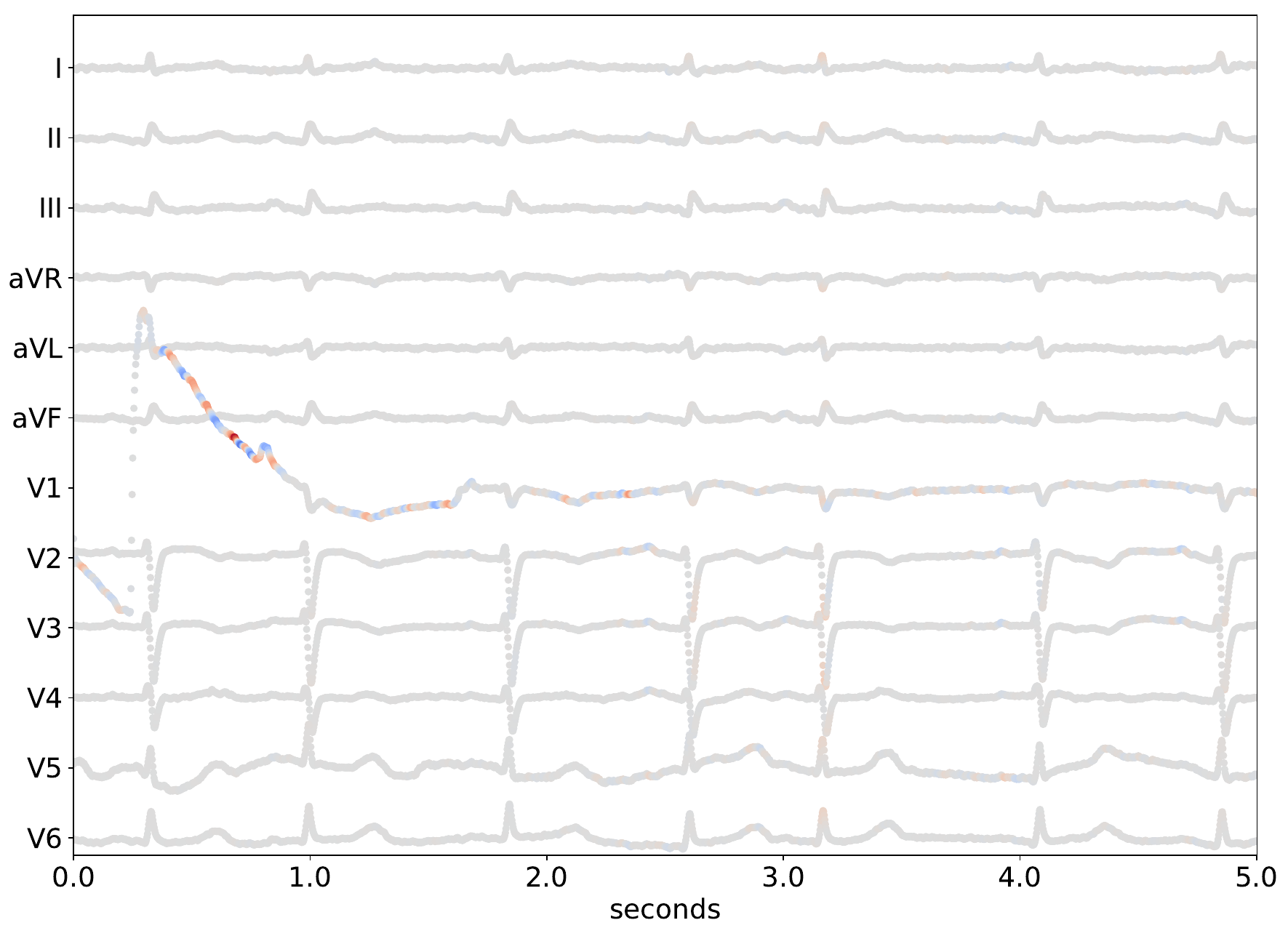}
    \caption{Positive (red) and negative (blue) relevance scores calculated with IG on false negative classified ECG (AF: $\sim0.008$) from CPSC data set (ID A1017). Relevance scores are clustered around the artefact in lead V1.\label{fig-relevances-1017}}
\end{figure}
\begin{figure*}[t!]
    \centering
	\includegraphics[width=0.99\textwidth]{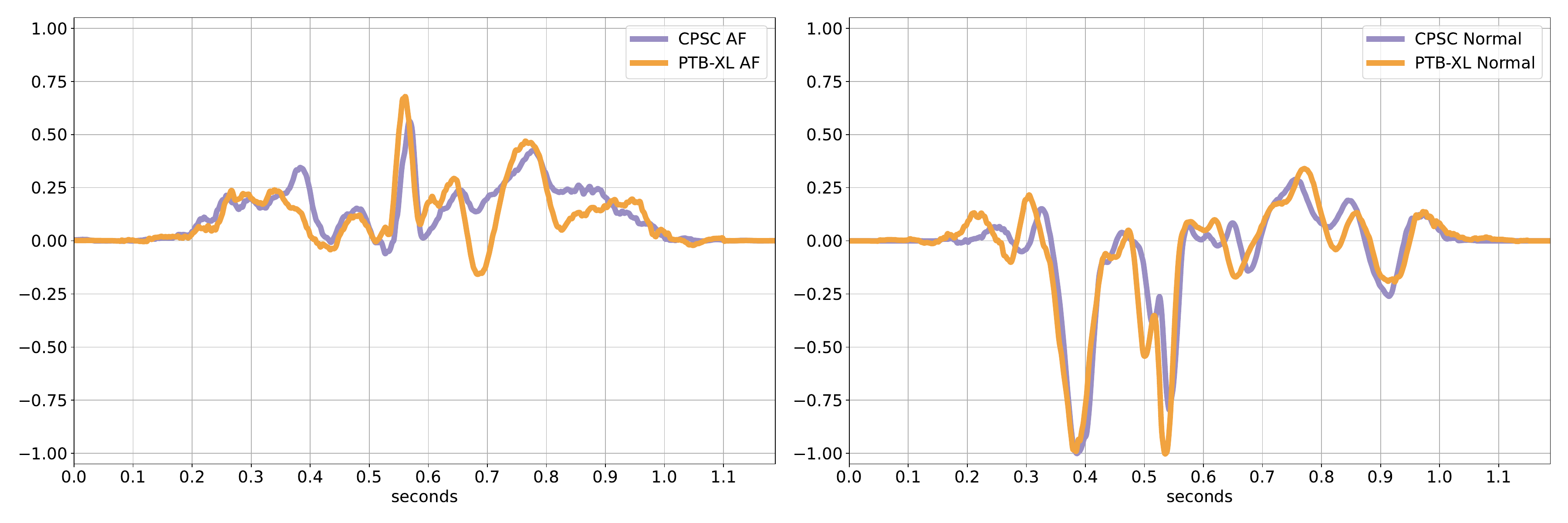}
	\caption{IG relevance scores for lead V1 averaged over $200$ ECGs extracted from CPSC (blue) and PTB-XL (orange). Figures depict AF recordings (left) and normal recordings (right), respectively.}
	\label{fig-avg-CPSCvsPTB}
\end{figure*}
\begin{figure}[!t]
    \centering
	\includegraphics[width=\columnwidth]{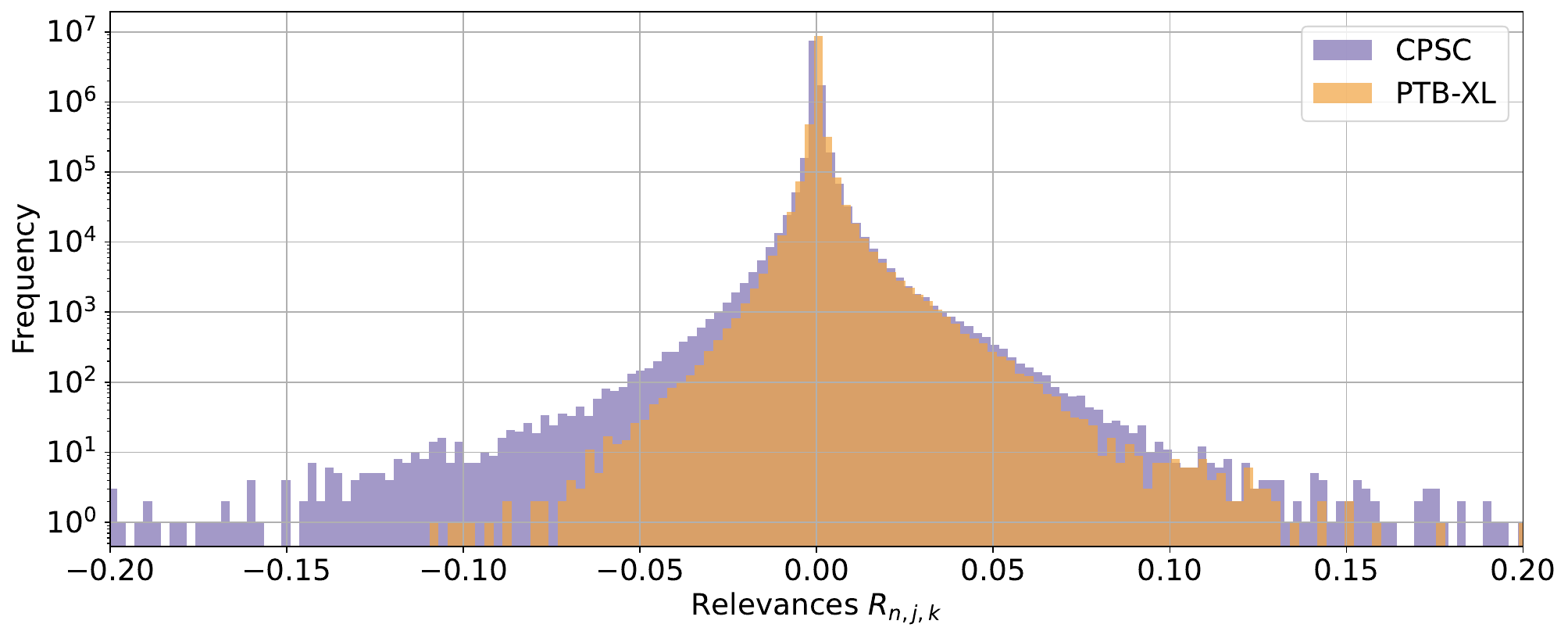}
	\caption{Relevance scores of LBBB recordings from CPSC database (blue) compared to PTB-XL data (orange). To increase visibility, the x-axis is limited to $[-0.20,0.20]$. Values for LBBB range from $[-0.11,0.21]$ and values for normal recordings from $[-0.64,0.56]$.}
    \label{fig-hist-LBBBvsSR-PTB}
\end{figure}
\begin{figure}[t!]
    \includegraphics[width=\columnwidth]{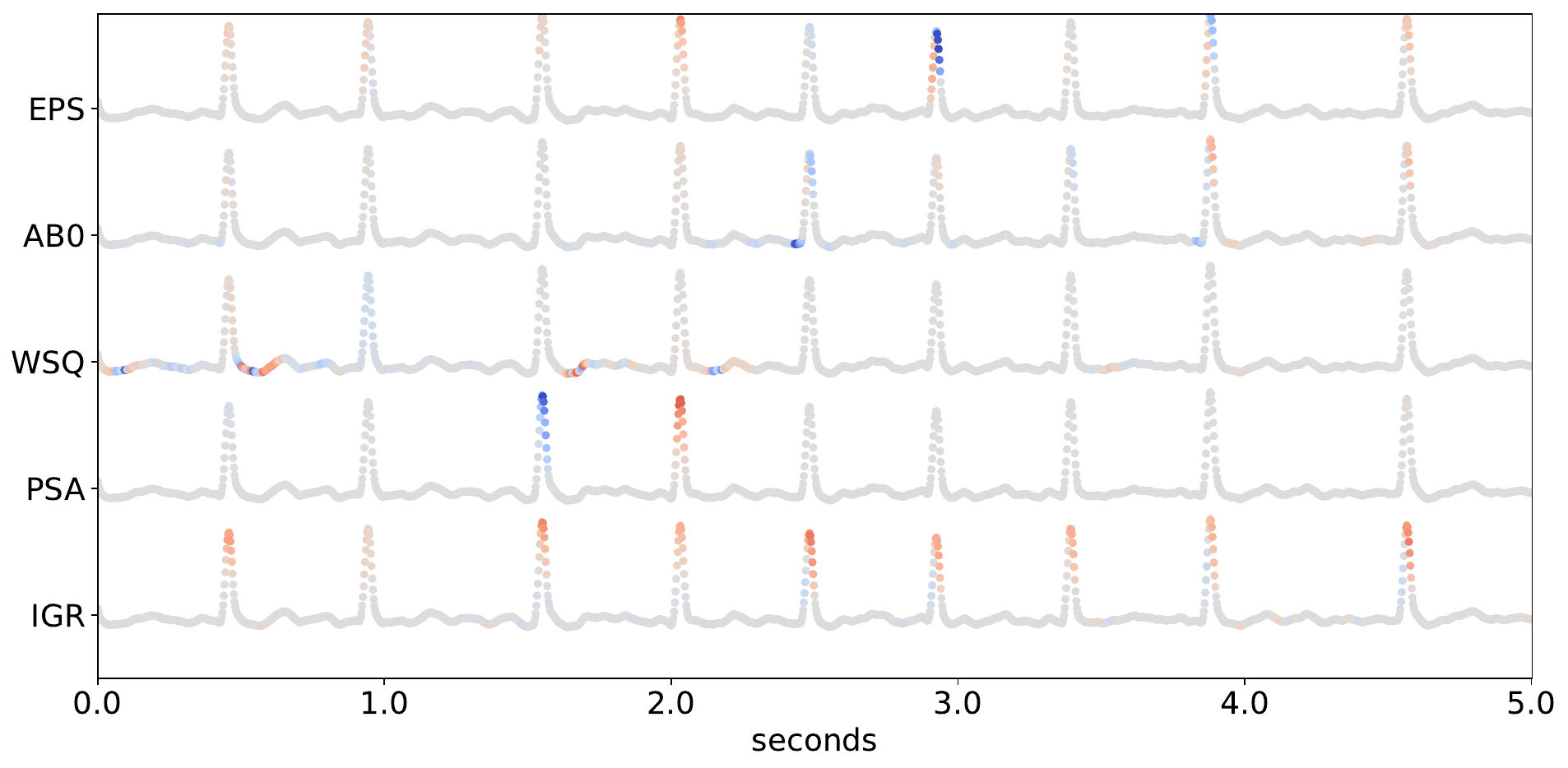}
    \caption{Relevance scores calculated with five XAI methods normed to $[-1,1]$ each on lead V6 of a correctly classified electrocardiogram (AF: $\sim0.987$) from CPSC data set (ID A0086). EPS: LRPEpsilon, AB0: LRPAlpha1Beta0, WSQ: LRPWSquare, PSA: LRPSequentialPresetA, IGR: IntegratedGradients. \label{fig-relevances-86-V6}}
\end{figure}

When looking at individual recordings we also observed that in cases of artefacts, such as baseline drifts or noise, IG relevance scores are usually accumulated mainly in these areas. This can be seen on multiple false negative classifications, such as recordings A1017 (lead V1, Fig. \ref{fig-relevances-1017}), A0745 (V6), and A0205, A0502 (both multiple leads, mainly: V1-6). In some cases the classification was still correct despite the focus on artefacts, e.g. A0639 (V1) classified as AF with $\approx0.904$.

\subsection{Comparison of Databases}
We repeated all experiments conducted on the CPSC database using data from PTB-XL instead. All quantitative methods show similar results for PTB-XL data, exemplarily shown for average beats in AF classification in Fig. \ref{fig-avg-CPSCvsPTB}. Especially the distribution of relevance scores for LBBB recordings is narrower and shifted closer to positive values than for CPSC data (Fig. \ref{fig-hist-LBBBvsSR-PTB}).

\subsection{Comparison of XAI Methods}
IG and all considered LRP methods yield diverging results for the given data set. As can be seen in Fig. \ref{fig-relevances-86-V6} as an example, LRP methods $\epsilon$ and $\alpha\beta$ distribute high absolute relevance scores especially around R-peaks, while $\omega^2$ shows higher absolute values on waves in between as well as artefacts. IG can also concentrate high absolute relevance scores around artefacts, but generally shows more high absolute values, especially on R peaks, when comparing leads of single patients to each other.

\section{Discussion}


Results of the first analysis show that IG relevance scores follow a reasonable distribution (Fig. \ref{fig-hist}) with the majority of values being close to zero. This is expected as the majority of samples in an ECG is at baseline, e.g. the interval between two heart beats from the end of the T-wave to the beginning of the P-wave, and carry little clinically-relevant information. Comparing AF and LBBB classification shows that the AF relevance scores are more evenly spread around zero while the LBBB relevance scores tend to more positive relevance scores which can also be seen clearly in Fig. \ref{fig-hist-LBBBvsSR} with a distinct gap for positive relevance scores between LBBB and normal recordings. We conclude that the DNN trained a larger inter-class distance for LBBB classification.

Analyzing individual recordings (Fig. \ref{fig-boxplotR}) shows similar distributions for both classifications. Additionally, a distinct relationship between the averaged relevance scores $M_{n}$ and the probability of the DNN $C_n$ can be observed. An optimal DNN classifier would show a cluster nearby $C_n = 0$ and $M_n \ll 0$ for normal recordings as well as a cluster nearby $C_n = 1$ and $M_n \gg 0$ for AF/LBBB. The analyzed DNN shows a sub-optimal relationship that can generally be expected with a transition area between both clusters in which the DNN does not have high certainty in its decisions (e.g. Fig. \ref{fig-boxplotR-AFvsSR}: $C_n \in [0.1,0.4]$). Furthermore, we observed many of the false negative classifications slightly below the threshold, indicating that the thresholds might not be optimal for the CPSC data set.

When analyzing individual leads, significant differences in relevance score distributions between abnormal and normal recordings were revealed (Fig. \ref{fig-boxplot}). This indicates which leads are most relevant for the DNNs decision. In general, for AF, the limb leads show lower relevance scores compared to the chest leads \cite{Bollmann.2006}. For AF as well as LBBB classifications, lead V1 shows clear positive relevance scores, indicating that the DNN trained clinically-relevant features: For AF, f-waves can often be observed in V1 \cite{langley.2000} and for LBBB a negative terminal deflection in V1, e.g. a rS-complex with a tiny R-wave and a huge S-wave, is a clear diagnostic marker \cite{strauss.2011}. 
Interestingly, there is a large difference in the distributions of the precordial leads V4-V6. While in AF it shows a clear tendency towards positive relevance scores, for LBBB the median is close to zero. Another sign for LBBB are prolonged R-waves and absence of Q-waves in left-sided leads \cite{macfarlane.2020} which might not have been learned.

For these first analyses, we used averaged mean values of relevance scores, which have been used for explanations of models that take feature based input instead of raw data \cite{Lauritsen.2020, Jansen.2019}. However, this is a rather coarse measure. As the relevance scores are signed, values can be composed of rather low relevance scores or competing strong relevance scores for and against the respective class. Still, outliers in overall means or means of leads could be an indicator for false classification due to artefacts, for example if a lead not typically being relevant for this abnormality has the highest mean, such as in lead V6 in Fig. \ref{fig-boxplot-LBBBvsSR}.

As time information is lost in average means, we proposed the third analysis. As can be seen in Figs. \ref{fig-avg-AFvsSR} and \ref{fig-avg-LBBBvsSR}, the ''average beat'' and ''average relevance scores'' of a single lead can give an even more detailed idea of the model's features. 
Although it is still not possible to uniquely identify the actual features learned by the DNN, positively relevant areas in case of missing P-waves for AF classification indicate a good fit to clinical criteria \cite{langley.2000}. Additionally, for the healthy controls, there are very pronounced negatively relevant areas nearby P-waves, demonstrating that the DNN learned that existence of P-waves is a counter-sign for AF. As IG does not allow to gain insight into the time scale, we cannot quantify to what extent RR-interval variations impact relevance scores. However, as the QRS-complex has similar shapes in AF and normal recordings, we assume that the DNN took the arrhythmic RR-intervals of AF recordings into account.

\begin{figure*}[!t]
    \centering
    \begin{subfigure}[b]{0.48\textwidth}
        \centering
        \includegraphics[width=1\textwidth]{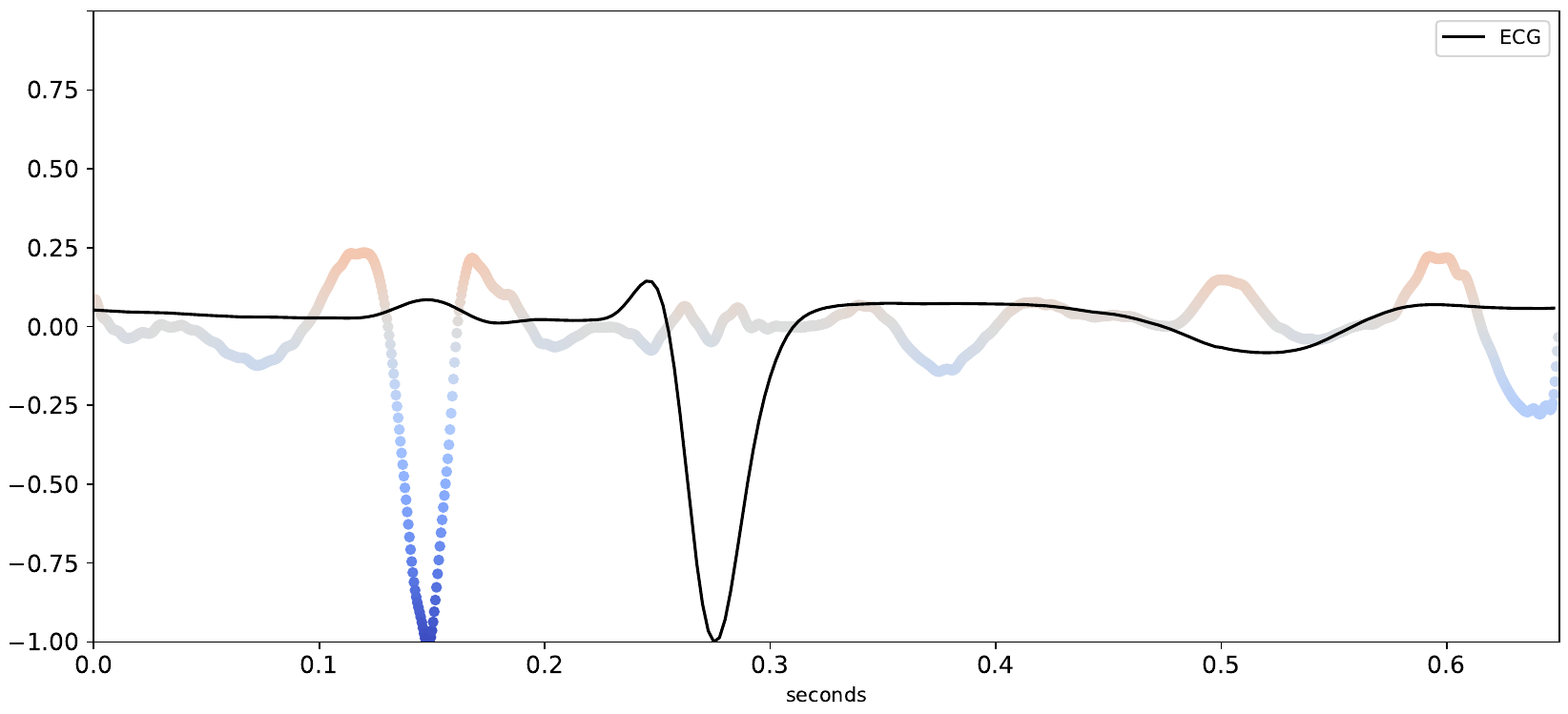}
    	\caption{Lead V1 used for AF classification.}
    	\label{fig-avg-single-AF}
    \end{subfigure}
    \hfill
    \begin{subfigure}[b]{0.48\textwidth}
        \centering
    	\includegraphics[width=1\textwidth]{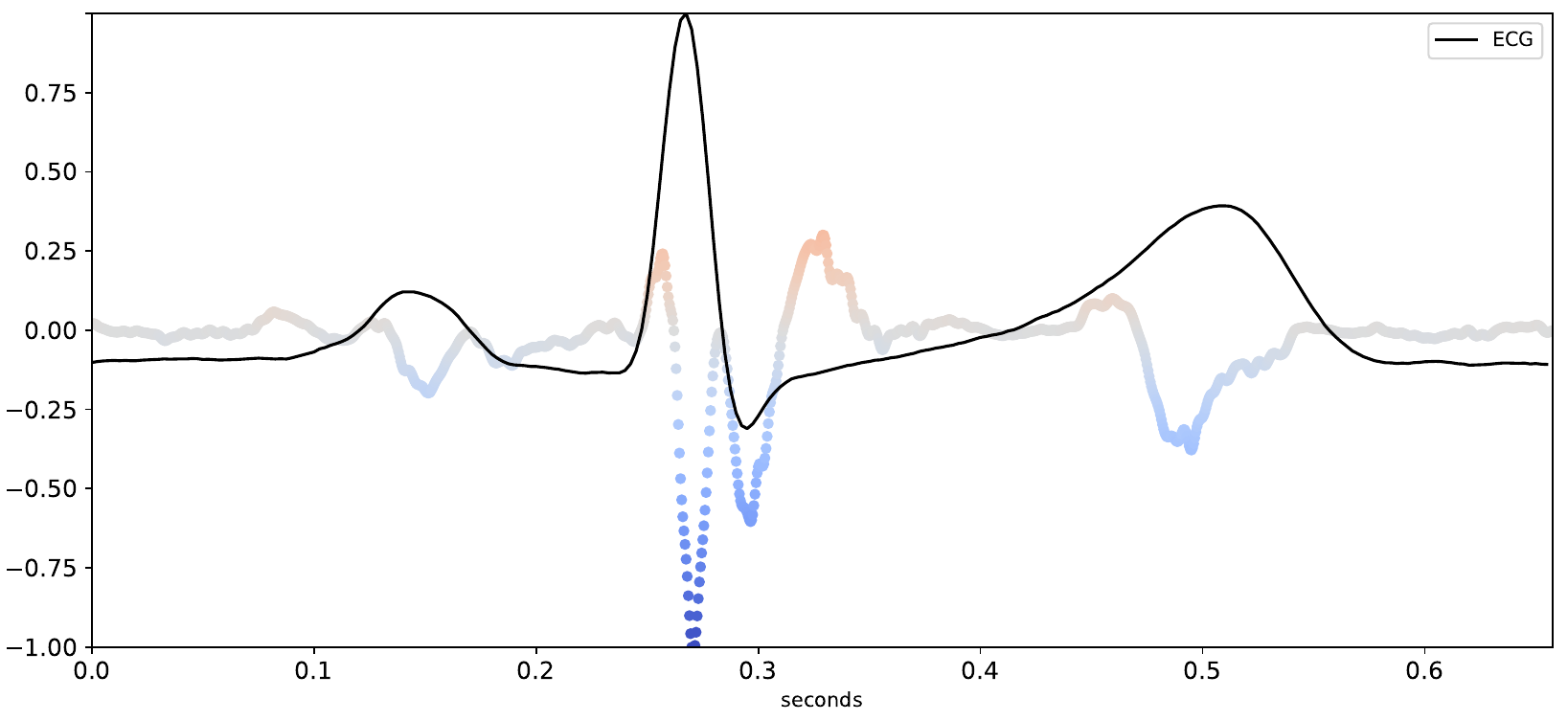}
    	\caption{Lead aVL used for LBBB classification.}
    	\label{fig-avg-single-LBBB}
    \end{subfigure}
    	\caption{Average beats (black curve) and relevance scores for individual leads in a single normal recording correctly classified by the DNN: a) Highly negative relevance scores (blue) are found during the occurrence of the P-wave. b) Negative relevance scores (blue) are found during the P-/T-waves, and especially during occurrence of the P-wave of the QRS-complex. }
\end{figure*}

Moreover, when analyzing the shape of an average relevance signal, which is continuously averaged over more and more recordings in Fig. \ref{fig-avg-AFvsSR} (see Supplemental Material for a video), it can be seen that, for AF as well as normal ECGs, the variance of relevance scores is quite low. This indicates a robustness of the DNN as it generates similar relevance scores despite the natural inter-patient variability in abnormal ECGs.
Regarding LBBB classification (Fig. \ref{fig-avg-LBBBvsSR}), high relevance scores around broadened QRS-complexes indicate a good fit to clinical criteria \cite{Tan.2020}. The criterion of a T-wave displacement opposite to the major deflection of the QRS-complex \cite{Tan.2020} can also be observed very well, although it results in small positive relevance scores only. In contrast, for healthy controls, T-waves result in very pronounced negatively relevant areas (e.g. Fig. \ref{fig-avg-single-LBBB}). Similarly, for AF classifications, P-waves are learned as a feature that indicates the absence of AF (e.g. Fig. \ref{fig-avg-single-AF}). Furthermore, the robustness of the relevance scores in terms of variance is even higher than for AF.

\subsubsection{Comparison of XAI methods}
In this work we applied the XAI attribution methods IG and LRP. There are other approaches available for explaining models for biosignal data using ante-hoc methods as in \cite{Hu.2021, Doborjeh.2021, elul.2021}, but these methods are not suitable for pre-trained DNNs where no adaption to the model itself is possible. Other methods, such as perturbation methods \cite{zeiler2014visualizing, zintgraf2017visualizing}, focus on occluding different parts of images and then analyzing the resulting changes in activations. These methods can also be used to calculate relevance scores for every input feature, but as shown by \cite{samek2015evaluating} they produce noisier heatmaps compared to LRP methods. Our results indicate that both methods, IG and LRP, are well suited for gaining insight into reasoning of DNNs applied to biosignals. Additionally, we conducted a comparison of IG and LRP methods (Fig. \ref{fig-relevances-86-V6}) and came to the conclusion that IG gives most distinct results.

\subsubsection{Comparison of databases}
To account for a change in the underlying data set, we validated our results on the CPSC database using PTB-XL instead and obtained similar results. One noticeable difference was observed in the relevance score distribution of LBBB recordings, where less negative values for PTB-XL could be explained by the more specific label ''Complete LBBB'', which might be easier to classify. These more differentiated labels bear the potential for comparison of model performance on complete and incomplete LBBBs.

\subsubsection{Artifacts}
We observed that the DNN tends to produce wrong classifications when artefacts are present as can be seen exemplarily in Fig. \ref{fig-relevances-1017}. This effect has been observed by others as well \cite{Taniguchi.2021}. Although we have not attempted it in this work, artefact detection based on our approach could be a promising avenue for future work. Additionally, we observed that the relevance scores result in certain temporal patterns that might allow the application of analysis methods from nonlinear signal processing \cite{apen} which we will analyze in future work.

\subsubsection{Key findings}
In summary, our analysis suggests that the model by Ribeiro et al. learned features similar to cardiology textbook knowledge. IG relevance scores indicate that it learned features pointing towards a disease, such as the abnormal QRS-complex in LBBB, while other features, such as the T-wave pointing in opposite direction, are not used for LBBB detection. Instead, the opposite of the feature, a T-wave pointing in expected direction, is used as a feature for detecting healthy ECGs. Our proposed analysis and visualization methods for relevance scores facilitate a rapid and effective assessment of the DNN's learned features and were confirmed by cardiologists.

\subsubsection{Limitations}
However, a limitation of our analysis based on IG is that we cannot infer any time-dependent information of the relevance scores. Especially for AF it is not clear whether e.g. the R-peaks are marked as relevant because of their morphology or their distance to one another. Therefore, we rate our results as more robust for LBBB as a morphological abnormality compared to AF as an arrhythmic and therefore time-dependent abnormality. 
Another limitation of our work is that we used public ECG databases which might introduce a certain bias. Therefore, using a data set from actual clinical practice on a cardiology ward or in emergency care might show different results. Thus, in future work, we will verify our results with more diverse data sources.

\section{Conclusion}

Missing explainability of ML methods for ECG analysis is a pressing issue preventing the dissemination of these methods in clinical practice. In this work we aimed enabling an objective justification of a DNN's decision by analyzing a state-of-the-art DNN for ECG classification with different XAI methods and data from different databases. Although this approach does not provide absolute certainty about the features learned by the DNN, it allows for inferring assumptions about its decision process. For example, our results reveal that the DNN learned that clearly-visible P-waves are a counter-sign for AF and T-waves pointing in same direction as the QRS-complex in particular leads are counter-signs for LBBB. Furthermore, decisions of the DNN for LBBB classification are based on unusual QRS-complexes. We conclude that the DNN learned cardiology textbook knowledge covering the whole cardiac cycle including P-wave, QRS-complex and T-wave. Moreover, we were able to explain false classifications due to transient noise which attracts the DNN's relevance scores, leading to relevant features being ignored. 

In future work, we will use the methods proposed in this work for developing an interactive tool for clinical practice which offers cardiologists an intuitive overview of the DNN's reasoning, supporting them in their decision whether to trust the DNN's classification, or not.

\appendices

\section*{Competing Interests}
The authors declare no competing interests.

\section*{Code Availability}
All source code developed in this work is publicly available on GitLab: \url{https://gitlab.gwdg.de/medinfpub/biosignal-processing-group/xai-ecg}, commit \#aed722d8.

\section*{Supplementary Material}
We provide plots of all quantitative analyses on PTB-XL in \textit{PTB\_analyses.pdf} and videos showing average beats and relevance scores for all CPSC data: \textit{beats\_V1\_AF.mp4} (AF classification, lead V1), \textit{beats\_AVL\_LBBB.mp4} (LBBB classification, lead aVL).

\bibliographystyle{IEEEtran}
\bibliography{template}{}

\end{document}